\documentstyle[11pt,aaspp4]{article}
\begin{document}
\newcommand{\msun}{M_{\odot}}
\newcommand{\zsun}{Z_{\odot}}
\newcommand{\kms}{\, {\rm Km\, s}^{-1}}
\newcommand{\cm}{\, {\rm cm}}
\newcommand{\mpc}{\, {\rm Mpc}}
\newcommand{\seg}{\, {\rm s}}
\newcommand{\hz}{\, {\rm Hz}}
\newcommand{\hi}{H\thinspace I\ }
\newcommand{\hii}{H\thinspace II\ }
\newcommand{\heii}{He\thinspace II\ }
\newcommand{\heiii}{He\thinspace III\ }
\newcommand{\nhi}{N_{HI}}
\newcommand{\lya}{Ly$\alpha$ }
\newcommand{\etal}{et al.\ }
\newcommand{\yr}{\, {\rm yr}}
\newcommand{\eq}{eq.\ }
\title{Reionization of the Intergalactic Medium and
the Damping Wing of the Gunn-Peterson Trough}
\author{Jordi Miralda-Escud\'e}
\affil{University of Pennsylvania, Dept. of Physics and Astronomy,
David Rittenhouse Lab.,
209 S. 33rd St., Philadelphia, PA 19104}
\authoremail{jordi@llull.physics.upenn.edu}

\begin{abstract}

  Observations of high-redshift quasars show that the IGM must have been
reionized at some redshift $z>5$. If a source of radiation could be
observed at the rest-frame \lya wavelength, at a sufficiently high
redshift where some of the IGM in the line-of-sight was not yet
reionized, the Gunn-Peterson trough should be present. Longward of the
\lya wavelength, a damping wing should be observed caused by the neutral
IGM whose absorption profile can be predicted. Measuring the shape of
this damping wing would provide irrefutable evidence of the observation
of the IGM before reionization, and a determination of the density of
the neutral IGM. This measurement might be hindered by the possible
presence of a dense absorption system associated with the source.

  Shortward of the \lya wavelength, absorption should be seen from the
patchy structure of the IGM in the process of reionization, intersected
in the line-of-sight. We show that a complete Gunn-Peterson trough is
most likely to continue to be observed through the epoch where the IGM
is partially ionized. The damping wings of the neutral patches should
overlap if the proper pathlength through an ionized region is less than
$1 h^{-1} \mpc$; even in larger ionized regions, the characteristic
background intensity should be low enough to yield a very high optical
depth due to the residual neutral fraction, although occasionally some
flux may be transmitted through large, underdense voids within an
ionized region. The case of the \heii reionization is also discussed,
and we argue that helium was already doubly ionized by $z=3$ throughout
the IGM.

  The recently discovered afterglows of gamma-ray bursts might soon be
observed at the very high redshifts required for these observations.
Their featureless continuum spectrum and high luminosities make them
ideal sources for studying absorption by the IGM.

\end{abstract}


\section{Introduction}

  Observations of the spectra of high-redshift quasars blueward of the
hydrogen \lya line have demonstrated so far that the universe was
already ionized by $z \simeq 5$, the highest redshift at which any
sources have been found so far (Schneider, Schmidt, \& Gunn 1991; Franx
\etal 1997). If the intergalactic medium (hereafter IGM) in the
line-of-sight to a source was neutral, the resonance scattering at the
wavelength of the \lya line should cause the ``Gunn-Peterson trough''
(Gunn \& Peterson 1965), with an extremely high optical depth that would
entirely absorb the flux from the source for any reasonable density of
the IGM. At the same time, we expect that the known sources of ionizing
photons (comprising quasars, stellar objects and hot gas) started to
form much later than the recombination epoch, unless the amplitude of
density fluctuations was much larger on small scales than predicted by
all the models of structure formation. This implies the IGM should have
been reionized at some point. In fact, the COBE measurement of the
electromagnetic spectrum of the microwave background (with a limit on
the $y$ parameter $y < 1.5\times 10^{-5}$; Fixsen \etal 1996), as well as
the presence of the Doppler peak in the intensity fluctuations, show
that the IGM was reionized at $z\lesssim 300$
(Wright \etal 1994; Hu \& Sugiyama 1994; Tegmark \& Silk 1995).

  It is generally expected that the reionization was caused by the
ionizing radiation from galaxies and active galactic nuclei. This
implies that the IGM should have been reionized inhomogeneously, with
every source ionizing first its immediate vicinity; the ionized bubbles
would then fill a growing fraction of the volume in the universe until
they overlap (Arons \& Wingert 1972). Models of structure formation
predict that reionization did not occur at extremely high redshift,
but most likely at $z \lesssim 30$, given the epoch when the collapse of
small-scale density fluctuations led to the formation of the
first galaxies where enough stars could form to reionize the IGM (e.g.,
Tegmark, Silk, \& Blanchard 1994; Haiman, Rees, \& Loeb 1996).
If a source could be seen at a
redshift higher than that when most of the IGM was reionized, we should
obviously see the Gunn-Peterson trough. And at redshifts were the
reionization is already complete, we observe the \lya forest caused by
the residual neutral fraction in the highly ionized IGM (in
photoionization equilibrium with the background radiation), with density
fluctuations originated in the gravitational collapse of gas in the
developping large-scale structure.

  Thus, the question that naturally arises here is: how does the
transition from one regime to the other take place? What should we
expect to see in a spectrum where the increasing optical depth of the
\lya forest with redshift turns into the Gunn-Peterson trough? This is
the question that shall be addressed in this paper. Meiksin \& Madau
(1993) speculated that a new class of high column density absorbers
might appear at redshifts higher than the epoch when the reionization
was completed, due to the neutral patches left in the IGM. The
possible observable signatures in the \lya spectra of sources seen
at the time when the IGM still had this patchy ionized structure will
be investigated in more detail here.

\section{The Red Damping Wing}

  We consider a source that is observed through a neutral IGM with
uniform density. If the photons scattered in the \lya resonance were
in a perfectly narrow line, we would expect to see a sudden drop in the
flux at the \lya wavelength at the redshift of the source. In reality,
the optical depth to \lya scattering is broadened both by the velocity
distribution of the atoms and the natural width of the line. When the
IGM is mostly neutral, the intrinsic line width is the dominant effect.
The resulting absorption profile was discussed in Miralda-Escud\'e \&
Rees (1997), but we reproduce this here for completeness.

Any fluid element along a velocity interval $dV$ along the line-of-sight
(corresponding to the Hubble expansion velocity over a spatial interval
$dx = dV/H(z)$) has its optical depth spread in the observed spectrum
according to
$\tau(V) = (\tau_0 dV/c) R_{\alpha}/\pi/[(V/c)^2 + R_{\alpha}^2]$.
Here, $\tau_0$ is the Gunn-Peterson optical depth of the neutral IGM
before broadening, and is given
by $\tau_0= 2.1\times 10^5\, [\Omega_b h (1-Y)/0.03]\,
[H_0 (1+z)^{3/2}/H(z)]\, [(1+z)/6]^{3/2} $, where $Y$ is the helium
abundance and $H(z)$ is the Hubble constant at redshift $z$. The IGM
is assumed to contain all the baryons, with density $\Omega_b$ in
units of the critical density. We have defined also $R_{\alpha} =
\Lambda/(4\pi \nu_{\alpha} ) = 2.02\times 10^{-8}$, where
$\Lambda =6.25\times 10^8 \seg^{-1}$ is the decay constant for
the \lya resonance, and $\nu_{\alpha} = 2.47\times 10^{15} \hz$ is the
frequency of the \lya line.
As a result of this broadening, the
observed absorption profile, at a wavelength separation $\Delta \lambda$
on the red side of the Gunn-Peterson trough, should be given by
\begin{equation}
 \tau(\Delta\lambda) = {\tau_0\, R_\alpha\over \pi } \,
\int_{\Delta\lambda/ \lambda}^{\infty}\,
{d (V/c)\over (V/c)^2 + R_{\alpha}^2 } ~ .
\end{equation}
Since $\tau_0 \gg 1$, the shape of this profile can only be
observed when $(\Delta\lambda) / \lambda \gg R_{\alpha}$, so we have
\begin{equation}
\tau (\Delta\lambda) =
\tau_0 R_{\alpha}/\pi \left( {\Delta\lambda\over \lambda} \right)^{-1} =
1.3\times 10^{-3}\, {\Omega_b h (1-Y)\over 0.03}\,
{H_0\, (1+z)^{3/2}\over H(z)}\, {(1+z)\over 6}^{3/2}
\left( {\Delta\lambda / \lambda} \right)^{-1} ~ .
\label{dampp}
\end{equation}
Notice that the damping wing of the Gunn-Peterson effect has the optical
depth falling as the inverse of the wavelength separation, as opposed
to the inverse-square law for the damped absorption lines caused by
high column density systems. The reason is that as we increase the
wavelength separation, a larger pathlength through the neutral IGM
contributes significantly to the optical depth, providing therefore
a larger effective column density.

  If a source is found at very high redshift showing the Gunn-Peterson
trough, the presence or absence of the damping wing with the predicted
absorption profile can be an
unambiguous test for the state of ionization of the intervening IGM.
The observation of the Gunn-Peterson trough alone does not prove that
a source is being observed behind neutral IGM, because the value of
$\tau_0$ is so large that a small residual neutral fraction in 
a reionized IGM can reduce the transmitted flux to undetectable levels.
If the damping wing is observed, an obvious immediate application will
be to infer the parameter $\tau_0$ from the observed width. At very
high redshifts, we can use the approximation $\Omega(z)\simeq 1$
independently of the value of $\Omega$ at present, so 
$H(z)\simeq H_0\, \Omega^{1/2} (1+z)^{3/2}$. Thus, by measuring
$\tau_0$ we can infer the quantity $\Omega_b h\, \Omega^{-1/2}$.

  There are two possible caveats in the measurement of this quantity.
The first is that if the source is observed at the epoch when the IGM is
being reionized, some regions along the line-of-sight should be mostly
ionized and some others should be neutral. If the Hubble velocity across
the \hii regions is small compared to the width of the damping wing, the
form of the damping profile should be essentially indistinguishable from
the case of the homogeneous medium (and in that case, the width of the
damping wing measures the space-averaged density of the fraction of the
IGM that is neutral), but if the velocity is comparable then there will
be random variations in the shape of the damping wing depending on the
distribution of neutral gas close to the source.
From equation \ref{dampp}, the width of the damping wing should be
$\sim 1500 \kms$, or $\sim 1 h^{-1} \mpc$, at $z=5$. Notice that if
there is only an ionized region around the source itself, with the
rest of the gas in the line-of-sight that contributes significantly
to the damping wing being all neutral, this does not affect the
measurement of $\tau_0$, but it only affects the redshift of the
neutral gas closest to the source, which can be obtained from a
two-parameter fit to the shape of the observed damping wing.

  The second possible problem is that the neutral IGM may not be
homogeneous. In particular, any source of light is likely to be within
a system that has gravitationally collapsed, and if the source has not
ionized the surrounding gas producing an expanding cosmological \hii
region, we should expect it to be surrounded by a halo of accreting
neutral gas. In that case, an additional column density of neutral
hydrogen is added, very close to the redshift of the source. One
could always try to circumvent the problem by adding the ``local''
column density as a third parameter to the fit to the damping wing
profile, but for a realistic accuracy of the spectrum the errors will
of course greatly increase as more parameters are added to the fit.

  As an example, we plot in Figure 1 the absorption profile on a source
at $z=9$ ({\it solid line}), for $\tau_0=4.3\times 10^5$. We
have used the exact expression for the profile of the damping wing
due to the neutral IGM, presented in the Appendix, where we do not
assume $(\Delta \lambda)/\lambda \ll 1$ (as in \eq \ref{dampp}),
and we include the effect of the cosmological evolution of the
Gunn-Peterson optical depth, and the exact expression for the cross
section of the \lya resonance line. The
correction on the optical depth relative to the approximation in \eq
\ref{dampp} is about 20\% when $\Delta\lambda / \lambda = 0.02$, and
reaches a factor of 2 at $\Delta\lambda/ \lambda = 0.06$; it will
therefore be important to use the exact expression in the Appendix for
practical applications. We also plot in Fig. 1 the result of adding to
the IGM optical depth an absorbing system with column density
$N_{HI} = 2\times 10^{20} \cm^{-2}$ at the source redshift $z=9$
({\it dotted line}). Finally, the dashed line shows again the IGM
optical depth without any additional absorbing system, but with the
parameters $z=9.009$, $\tau_0=4.9\times 10^5$, which provide the
best fit to the dotted line. It is clear from this example that even
with an accuracy of 1\% in the determination of the shape of the
damping wing, there is still a large error in the determination of
$\tau_0$ if an absorber of unknown column density is present at
the redshift of the source. However, one may be able to determine the
source redshift independently (for example, from associated metal-line
absorption), in which case the ambiguity between the hydrogen column
density of an associated absorber and $\tau_0$ should be much
reduced.

\section{Flux Transmission on the Blue Side of the Gunn-Peterson Trough}

  Moving down in redshift from the source, a line-of-sight will
characteristically traverse several cosmological \hii regions and
neutral patches over the redshift interval corresponding to the epoch
when the IGM was being reionized. The process of reionization can be
very complicated: the sources of ionizing photons may have a wide range
of luminosities and may be short-lived, so some ionized regions will
start recombining if their source fades, becoming partially neutral
until they are ionized again by another source. The sources could be
highly clustered, so the ionized regions may start being very small and
due to faint, individual sources, and later grow to a much larger scale
and be ionized by clusters of sources, while large neutral patches may
still remain in the IGM. And of course, the \hii regions expand into a
highly inhomogeneous IGM which is gravitationally collapsing into new
large-scale structures. The case of a source of constant
luminosity surrounded by a homogeneous medium can be solved
analytically (Shapiro \& Giroux 1987).

  An ionized region intercepted by the line-of-sight over a total length
$L_i \equiv V_i/H(z)$ may cause a ``gap'' in the Gunn-Peterson trough if
the damping wings of two neutral patches on each side do not completely
overlap. Taking the parameters used in Fig. 1, we see that if we require
that at least a fraction $e^{-1}= 0.36$ of the flux be transmitted in
the middle of the gap, then the length of the intersected ionized region
must be at least $V_i/c > 10^{-2}$ (because at a separation $V/c = 5
\times 10^{-3}$ from the edge of a neutral zone, the optical depth of
the damping wing in Fig. 1 is $\tau=0.5$). Since $\tau_0 \propto
(1+z)^{3/2}$, this minimum size scales with redshift as $V_i/c >
10^{-2}\, [(1+z)/10]^{3/2}$ in the approximation of \eq \ref{dampp}, or
a proper length $L_i > 1 h^{-1} \mpc$ (these scales are also proportional
to $\Omega_b h (1-Y)$, but we are setting this quantity to $0.03$ in
this section).

  In order to observe any transmitted flux to the blue of the
Gunn-Peterson trough on a source at redshift $z_s$, some \hii region of
this size has to be present on the line-of-sight at a redshift $1+z_i >
1+z_{\beta} \equiv (1+z_s)/1.18$, since at lower redshifts the Ly$\beta$
Gunn-Peterson trough blocks the flux anyway. If the last neutral patch
in the line-of-sight is located at $z_n>z_{\beta}$, then flux may be
observable between the blue \lya damping wing due to this last neutral
patch and the Ly$\beta$ damping wing of the neutral IGM close to the
source, and some additional small gaps may be observable due to large
\hii regions at $z > z_n$. If $z_n < z_{\beta}$, then only these gaps due
to isolated \hii regions may be observable.

  The existence of ionized regions with a proper size larger than
$1 h^{-1}\mpc$ for a substantial period of time before reionization was
over depends, of course, on the type of sources causing the ionization.
Large \hii regions would be expected if the radiation was dominated by
very luminous sources; the luminosity required to reach a proper size
$1 h^{-1} \mpc$ at $z=10$ is similar to that of the most luminous
quasars known. Reionization by early galaxies has been considered more
likely in the usual hierarchical models for structure formation
(Tegmark, Silk, \& Blanchard 1994, Haiman \& Loeb 1997, MR97).
In this case, large \hii
regions might still exist if the sources were highly clustered,
probably in the sites of formation of the future galaxy clusters. Early
galaxies might in fact be clustered on large scales due to a high
biasing factor, which is expected if galaxies form in density peaks and
when the power spectrum has a slope close to $n=-3$, as is generally
predicted to be the case on small scales (Bardeen \etal 1986). On the
other hand, if reionization is caused by low-luminosity sources which
are not highly clustered on a proper scale of $1 h^{-1} \mpc$, then by
the time the ionized regions had overlapped to this scale there should
be very few neutral patches remaining in the IGM, (except for the
neutral regions that are dense and self-shielded, i.e., regions which
are neutral because of their high recombination rate, rather than the
lack of sources in their vicinity).

  Whether any flux is actually seen on the blue side of the damping
wing of the last neutral patch in a line-of-sight, or in any gap due
to an isolated ionized region, depends of course on the neutral
fraction in the ionized medium being low enough. The neutral fraction
should be in ionization equilibrium with the radiation field from the
sources in the \hii region, and will cause a ``\lya forest'', as
observed up to $z\lesssim 5$ so far;
even a very small neutral fraction can completely absorb any remaining
flux that is left between the damping wings of the neutral patches.

  To calculate the absorption in the ionized regions, let us
parameterize the mean volume emissivity of ionizing photons as
\begin{equation}
\epsilon = {n_e\over t}\, N_r ~,
\label{emdef}
\end{equation}
where $n_e$ is the mean electron density in the ionized IGM and $t$ is
the age of the universe. Thus, when $N_r=1$ the emissivity is such that
one ionizing photon will be emitted for each electron over the age of
the universe. Obviously, at the end of reionization the factor $N_r$
should be greater than unity, both because the comoving emissivity
should increase with time (and therefore the integrated number of
emitted photons is less than $\epsilon t$), and because electrons can
recombine several times during reionization, increasing the number of
photons that need to be emitted. The photon intensity is typically
$\epsilon\, L_i/(8\pi)$ (where we take the typical mean-free-path of
photons to be $\sim L_i/2$), and the implied photoionization rate is
$\Gamma = \epsilon\, L_i\, \bar\sigma/2$, where $\bar\sigma$ is the
photoionization cross section averaged over all emitted photons; for a
typical power-law slope $F_{\nu}\propto \nu^{-1.5}$ near the hydrogen
ionization edge, $\bar\sigma \simeq 2\times 10^{-18}\cm^2$. In an \hii
region having the minimum size that allows flux to be transmitted
(using as before the condition that the optical depth between the two
damping wings reaches a value less than unity), the length is $L_i =
[c/H(z)] (4 \tau_0 R_{\alpha}/\pi)$ (see \eq \ref{dampp}).
The neutral fraction in a region of average density is equal to
$\alpha n_e/\Gamma$, and the optical depth in the ionized region due
to this neutral fraction is:
\begin{equation}
\tau_i = \tau_0\, {\alpha\, n_e\over (n_e/t)\, N_r\,
(\tau_0 R_{\alpha}/\pi)\, [c/H(z)]\, \bar\sigma } =
{\pi \alpha \over 3 c \bar\sigma R_{\alpha} \, N_r } =
{ 360\, T_4^{-0.7} \over N_r} ~.
\end{equation}
Here, $T_4$ is the temperature of the ionized gas in units of $10^4$ K;
the temperature dependence is due to the recombination coefficient.
Notice that $\tau_i$ is the optical depth in the \hii region if the gas
density is equal to the mean baryonic density. In reality, the gas
density should of course fluctuate and the optical depth should
consequently vary with wavelength over the interval of \lya absorption
of the \hii region.

  Thus, we have found the result that the optical depth to \lya
scattering in an ionized region that is just large enough to emerge from
the damping wings of the Gunn-Peterson trough is independent of redshift
and of any cosmological parameters, and depends only on the
dimensionless number $N_r$, defined in \eq \ref{emdef}.

  Let us explore more carefully the interpretation of $N_r$. There are
two different cases that need to be considered. In case I, the sources
of ionizing photons turn on quickly enough so that recombinations are
negligible compared to the number of photons that need to be emitted.
Therefore, the IGM is reionized when one ionizing photon has been
emitted for each electron, and $N_r=\epsilon / \bar\epsilon$, where
$\bar\epsilon$ is the time-averaged comoving emissivity. Unless the
sources are highly synchronized, $N_r$ will not be larger than $\sim$ a
few. The other possibility, case II, is that reionization is limited by
the rate at which electrons can recombine. In that case, $N_r$ is simply
the mean number of recombinations for each electron that take place over
the age of the universe, $t$: if the sources turn on over a time $t_r$,
each electron undergoes $N_r (t_r/t)$ recombinations during the time
$t_r$, and the required emissivity is $\epsilon = (n_e/t_r) N_r (t_r/t)
= (n_e/t) N_r$. This mean number of recombinations is
\begin{equation}
 N_r = 0.91 \, {0.03\over \Omega_b h} \,
\left({1+z\over 10} \right)^{3/2}\,
{ < \! n_e^2 \! > \over < \! n_e \! > }  ~,
\label{nreq}
\end{equation}
where the last term is the electron clumping factor. Thus, at a fixed
redshift, the only way for $N_r$ to be large is that the IGM be highly
clumped, and the
clumps need to have a large covering factor through a line-of-sight in
an ionized region in order to absorb the photons effectively (if the
covering factor is small, the condition that the mean recombination rate
is equal to the mean photon emission rate implies that the clumps are
mostly neutral, therefore not increasing the electron clumping factor).
But if gas clumps with a high covering factor are responsible for the
recombinations and, therefore, for the absorption of most photons before
they reach the edge of the \hii region and contribute to increase its
size, then our estimate of the photoionization rate in the \hii region
is not valid because we assumed the mean-free-path of the ionizing
photons to be $L_i/2$, and it should now be reduced by the covering
factor of the clumps. As long as the recombinations do not take place
mostly at the edges of the \hii regions (due to a higher gas density
there), $N_r$ cannot be larger than a few even for a very clumpy IGM.
Notice that $N_r$ is very large when $z \gg 10$, but of course the
required size of an \hii region $L_i$ to allow for flux transmission
between the two damping wings becomes more implausibly large as the
redshift increases.

  We therefore conclude that the optical depth $\tau_i$ of an ionized
region before reionization is complete should generally not be smaller
than $\sim 100$. Rare exceptions to this are nevertheless allowed,
because the photoionization rate inside the \hii regions that we
estimate above is only a mean value; for example, a line-of-sight may
pass very close to a very luminous source. At the same time, the optical
depth $\tau_i$ is the value expected when the IGM density is equal to
the mean baryonic density of the universe, but in reality a `` \lya
forest'' should be caused by the IGM density fluctuations, just as
observed at lower redshift. In photoionization equilibrium, the optical
depth observed at a given wavelength is proportional to the square of
the gas density. Therefore, the optical depth should be reduced by a
factor of $100$ in a void that is underdense by a factor of 10, allowing
for some flux to be transmitted. The transmission of any flux when
$\tau_i$ is so large requires a void that is not only sufficiently
underdense, but also sufficiently large to prevent the thermal
broadening of the absorption due to the gas in adjacent structures, with
typical overdensities near unity and temperature $\sim 10^4$ K, from
overlapping and obstructing the window. For example, for absorbers with
central optical depth $\sim 100$ and velocity dispersion $\sigma=10
\kms$ the thermal wings should extend to about $3\sigma$ on each side,
requiring the void to have a diameter of at least $\sim (60 \kms)/H(z)$.

  Possibly, the epoch of partial ionization in the IGM may be better
observed in the Ly$\beta$ line (and higher order Lyman series), where
the optical depth is reduced everywhere by a factor $0.16$ relative to
\lya. This could work in a source observed at a redshift close to the
end of the reionization epoch, and if the intensity of the ionizing
background rises quickly enough that, by the time the scale factor of
the universe has increased by $1.18$, significant flux is already being
transmitted through the \lya forest. But of course, contamination by the
\lya forest will make the analysis of the Ly$\beta$ spectrum very
complex.

  To summarize, the picture that emerges from this discussion is the
following: at the redshift $z_n$ of the last neutral patch of the IGM on
the line-of-sight from the source, a blue damping wing of absorption
should be produced by the neutral gas. However, the ionized IGM at
redshifts immediately below $z_n$ should normally have a large enough
neutral fraction, in equilibrium with the ambient intensity of ionizing
photons, to completely obstruct the flux from the source at the \lya
resonance, thus preventing the observation of the shape of the blue
damping wing. The highest redshift at which transmitted flux will be
seen should be that of a sufficiently underdense and large void, which
can open up a ``window'' in the saturated absorption of the ionized IGM.
Some of these voids might be located in the region where the optical
depth of the blue damping wing is still significant, or even between two
damping wings encircling an ionized region at $z> z_n$ with a size
larger than $L_i$.

  At $z < z_n$, the intensity of the ionizing background, $J_{HI}$,
should rise as the mean-free-path of the photons increases, leading to
an increase in the number of ``windows'' of transmitted flux associated
with voids. The redshift evolution of these ``windows'' will provide
information on the evolution of $J_{HI}$. How fast should the rise of
$J_{HI}$ be? This depends on the rate at which the ionized material can
recombine. In case I (where the mean rate of recombinations is much
lower than the mean rate of emissions before the \hii regions overlap),
reionization occurs after one ionizing photon has been emitted for
every proton in the IGM. At this point, the photon mean-free-path should
rapidly increase by a large factor starting from the typical size of the
\hii regions, thereby causing a fast increase in the fraction of
transmitted flux in \lya spectra. The intensity of the background rises
proportionally to $t-t_i$, where $t_i$ is the time at the end of
reionization. The increased background radiation penetrates regions of
denser gas, increasing the global rate of recombinations in the universe
until they can balance the emission rate. This stops the linear growth
of $J_{HI}$ with time, and the photon mean-free-path has grown to the
mean separation along a random line-of-sight between the Lyman limit
systems that have now been ionized, where enough recombinations take
place to match the mean emission rate. The subsequent evolution is known
observationally: at $z\simeq 4$ to $5$ the abundance of Lyman limit
systems is still large enough to be the main factor limiting the
intensity $J_{HI}$, and by $z \lesssim 2$ the mean-free-path has grown
to the horizon scale, and $J_{HI}$ is limited mainly by the redshifting
of photons.

  On the other hand, in case II recombinations are already important
before the \hii regions start to overlap. At this time, the
mean-free-path between Lyman limit systems associated with dense,
self-shielded structures is about the same as the typical size of the
\hii regions. Thus, there is no sudden increase of the mean-free-path
when the \hii regions overlap. As the emission rate of photons increases
and the universe expands, the self-shielded regions shrink in size and
the newly ionized dense gas enables the global recombination rate to
keep up with the emission rate, and this continues into the observed
epoch at $z<5$ as discussed previously. This leads to a much more
gradual increase of the mean-free-path and of $J_{HI}$ compared to case
I. Thus, the width of the redshift interval over which the \lya forest
increases its opacity, leading to the Gunn-Peterson trough, should give
us information on the clumpiness in the IGM, which determines the rate
of recombination at various stages of the reionization.

\section{The \heii Gunn-Peterson Trough}

  The first observations of absorption by \heii at $\lambda = 304
{\rm \AA}$ have been performed over the last few years (Jakobsen \etal
1994; Tytler \etal 1995; Davidsen, Kriss, \& Zheng 1996;
Hogan, Anderson, \& Rugers 1997; Reimers \etal 1997).
The observation by Jakobsen \etal (1994) was consistent with a complete
``Gunn-Peterson trough'', although the signal-to-noise that could be
achieved allowed them only to place an upper limit to the mean
transmitted flux of 20\% of the total at $z\simeq 3.2$. The \heii
\lya optical depth from a uniform IGM where all the helium is only once
ionized is
\begin{equation}
\tau_{0, \heii} = 2.3\times 10^3\, (\Omega_b h Y / 0.01)\,
[ H_0 (1+z)^{3/2}/H(z) ] \, [(1+z)/4]^{3/2} ~. 
\label{t0heii}
\end{equation}
Therefore, the observation of Jakobsen \etal could imply that the helium
in the IGM was not yet doubly ionized, but it was also consistent with
a typical \heii fraction much smaller than unity (this remains true in a
realistic model of an inhomogeneous IGM with on-going structure
formation).

  Davidsen \etal (1996) were the first to find evidence for the presence
of transmitted flux to the blue of the \heii \lya line in another
quasar, with a fraction $0.36 \pm 0.03$ over the redshift range $2.2 < z
< 2.7$. The detection of flux immediately proves that the helium in the
IGM was mostly twice ionized at $z < 2.7$, although it is possible that
some patches of \heii still remained at this epoch if the double
reionization of helium was not yet complete. In fact, Reimers \etal
(1997, hereafter R97) detected transmitted flux with much higher
resolution in a second quasar, QSO HE 2347-4342, in the range $2.8 < z
< 2.9$. All the transmitted flux appears in two windows, one at
$\lambda = 1160 {\rm \AA}$ of width $4 {rm \AA}$, and another at
$\lambda = 1174 {\rm \AA}$ of width $2 {\rm \AA}$. The corresponding \hi
\lya absorption spectrum (see Fig. 4 in R97) shows some weak absorption
features in the wider window, and no detectable absorption in the
narrower one. In particular,
\hi absorption at $\lambda = 1159$ \AA, in the middle of the
wider \heii window, is clearly detected. On the other hand, there are
other regions free of any detectable absorption in \hi (e.g., at
$\lambda= 1171$ \AA) where the flux is completely absorbed in \heii.
This demonstrates that the various regions in the IGM observed in this
line-of-sight cannot all be photoionized by an ionizing background with
approximately the same intensity, as shown in detail by R97 (see their
Fig. 5). It is clear that, while in the two windows of the \heii
spectrum the helium in the IGM must be mostly twice ionized, the other
regions with low \hi absorption must either have most of their helium
only once ionized, or must have a much lower intensity of \heii ionizing
photons.

  The double ionization of helium could very well occur at a later epoch
than the hydrogen ionization. If recombinations are not important during
the reionization (case I as defined in the previous section), \heii will
be ionized later if the ratio $I_{\heii}/I_{\hi}$ of emitted photons
above the \heii ionization edge and above the \hi ionization edge is
lower than $0.08$. But if many recombinations take place (case II), then
we only need $I_{\heii}/I_{\hi} < 0.5$, because \heii recombines 5.5
times faster than \hi (Miralda-Escud\'e \& Rees 1994). If quasars or
starburst galaxies produce the ionizing background, then
$I_{\heii}/I_{\hi} \lesssim 0.15$ (Haardt \& Madau 1996 and references
therein). Since even in the absence of clumpiness, the \heii
recombination time is equal to the Hubble time at $z\simeq 3$, a
delayed reionization of \heii relative to \hi is clearly not just
possible, but very likely.

  This suggests that we could be observing the kind of patchy helium
medium that we discussed previously for the case of hydrogen, and that
the two windows of transmitted flux observed by R97 could be \heiii
regions surrounded by \heii patches. However, the story for the \heii
ionization is quite different from the case of hydrogen. The difference
lies in the much larger thickness of the ionization fronts in the case
of helium, due to the lower helium abundance by a factor $0.08$, and the
lower cross section of \heii by a factor 4. At $z=3$ and in a region of
density equal to the mean, a \heiii ionization front should have a
proper thickness $\sim 1 \mpc$, comparable to the largest \heiii regions
that could be ionized by a single quasar. The corresponding velocity
width is $\sim 500 \kms$ (much larger than the expected width of the
damping wing for the \heii Gunn-Peterson trough, which is smaller than
that of \hi by a factor 0.08 at the same redshift).

  The large thickness of the \heii ionization fronts implies that,
during the epoch when the low-density IGM still contains a high fraction
of \heii, it is not likely that any windows of transmitted flux can be
seen through the \heii Gunn-Peterson absorption. In fact, if we repeat
the analysis of the previous section to obtain the optical depth
$\tau_{i,\heii}$ of a \heiii region due to the \heii remaining in
photoionization equilibrium, but this time expressing the answer in
terms of the length of the \heiii region $L_i = V_i/H(z)$, we obtain
\begin{equation}
\tau_{i,\heii} = { \tau_{0,\heii} \over N_r }\, {750 \kms \over V_i} ~.
\end{equation}
Since the windows of transmitted flux observed in R97 have velocity
widths similar to $V_i = 750 \kms$, and $N_r$ cannot be a large number
due to the same arguments used in the last section, we conclude that the
optical depth in a \heiii region would in fact be close to
$\tau_{0,\heii}$ (\eq \ref{t0heii}) if the double reionization of helium
was not yet complete. Another simple way to understand this result is
that when the thickness of the ionization front is similar to the size
of the \heiii regions, the fraction of \heii in the \heiii regions is
not much less than unity.

  The fact that the Davidsen \etal observations can be fitted with
models where the helium is already doubly ionized at $z=2.7$, with the
expected shape of the spectrum for the ionizing background when quasars
are the dominant sources (Miralda-Escud\'e \etal 1996, Croft \etal 1997)
also suggests that the \heii reionization should be complete, and the
state of the IGM should not be radically different at $z=2.8$. The
obvious differences in the \heii to \hi column densities presented by
R97 can be explained by large fluctuations in the intensity of the
background above the \heii ionization edge, which are expected due to
the large number of optically thick absorbers in \heii, but they do not
demand a large \heii fraction in the regions free of \hi absorption with
no transmitted flux at \heii. As an example, consider a typical void in
the IGM at $z\sim 3$ with a gas density a few times below average, where
the hydrogen neutral fraction is $\sim 10^{-6}$. This yields an optical
depth in the \hi \lya spectrum of $\sim 2\%$ (e.g., Croft \etal 1997),
which would not be detected in any of the regions that are apparently
free of absorption in the \hi spectrum in Fig. 4 of R97 due to the
uncertain continuum level. Given the mean spectrum of the ionizing 
ackground obtained from quasar models (Haardt \& Madau 1996), with
$J_{\hi}/J_{\heii} \simeq 50$, the fraction of \heii in the same void
should be $10^{-3}$, yielding a \heii \lya optical depth of $0.4$, which
could perfectly well be the case for the two windows of flux in R97. But
in another similar void where $J_{\hi}/J_{\heii} = 500$ (due to
absorption of the nearest quasars), the \heii optical depth should be
$4$ while the \heii fraction is still as low as $10^{-2}$.

\section{Conclusions}

  In this paper we have attempted to predict the observable features
that should be present in the spectrum of a high-redshift source seen
behind a region of neutral IGM, still to be reionized.

  The first feature is the damping wing of the Gunn-Peterson trough on
the red side of the \lya line. The width of the absorption profile of
this damping wing provides a method to measure the density of the
neutral IGM near the redshift of the source. There are two main
difficulties for this measurement: the possibility of an absorption
system large enough to contaminate the profile of the damping wing, and
uncertainties in the intrinsic emission spectrum of the source. The
second difficulty is particularly severe if the emitting source is a
quasar, since quasars have strong, broad \lya emission lines with
profiles that are highly variable. The second feature we have discussed
is the transition from the Gunn-Peterson trough to the \lya forest. We
have found that the \lya forest should gradually become thicker with
increasing redshift as the epoch of reionization is approached; at
redshifts where the IGM was only partially ionized, the presence of any
transmitted flux in individual cosmological \hii regions surrounded by
neutral patches of the IGM should be rare, due to the damping wings of
the neutral zones and the very high \lya optical depth that is still
caused by \hii regions with the characteristic intensity of the ambient
ionizing background that should be expected. It is also possible that
some transmitted flux could be seen from \hii regions in Ly$\beta$,
although that would of course be contaminated by the \lya forest.

  What type of sources can we expect to discover at ever higher
redshifts in the future, to finally be able to probe the epoch of
reionization? Quasars are the most luminous known sources, but there are
signs that their abundance is declining at $ z \gtrsim 3$ (Warren,
Hewett, \& Osmer 1994; Schmidt, Schneider, \& Gunn 1995) and it is not
clear if they can be found at much higher redshifts. Young galaxies must
necessarily have existed when the IGM was partially ionized if they are
the sources that caused the reionization, but the expectation is that
they are hopelessly faint for obtaining good quality spectra (Haiman \&
Loeb 1997; MR97). Recently, a new class of luminous extragalactic
sources in the ultraviolet has been identified: the afterglows of
gamma-ray bursts (Costa \etal 1997a; van Paradijs \etal 1997). The
afterglow was predicted from the cosmological fireball model
(M\'esz\'aros \& Rees 1997; Vietri 1997), and found to be in good
agreement with observations (e.g., Wijers, Rees, \& M\'esz\'aros 1997a;
Waxman 1997). The third observation of an afterglow (Costa \etal 1997b;
Djorgovski \etal 1997) resulted in the detection of an absorption system
in the spectrum at $z=0.83$ (Metzger \etal 1997), which has definitely
put to rest any doubts on the cosmological nature of GRBs. The optical
magnitudes of the GRB afterglows are similar to the luminous
high-redshift quasars. It is quite plausible that these afterglows will
be found in the coming years at redshifts higher than any other known
sources; in fact, if GRBs are produced in objects belonging to young
stellar populations, then they should take place soon after the first
stars were formed in the universe (Wijers \etal 1997b). At very high
redshifts, the probability of gravitational lensing is high, and this
can help in boosting the apparent brightness of some GRBs and their
afterglows. The very important advantage of GRB afterglows is that their
intrinsic spectrum is supposed to be a featureless power-law on small
wavelength ranges, because the radiation originates from synchrotron
emission in a relativistic shock. This eliminates the severe systematic
error of measuring the absorption profile of a damping wing near an
emission line in the case of a quasar. Thus, unless gamma-ray bursts did
not take place for some reason until a much later time than the epoch
when the first stars (or the first sources that caused the reionization)
formed, their afterglows should become a precious tool to study the IGM
at very high redshift.

\acknowledgements

  I am grateful to Martin Rees for numerous discussions that resulted
in many of the ideas presented here. I also thank Peter M\'esz\'aros
and Eli Waxman for discussions on gamma-ray bursts.

\newpage
\appendix{{\bf Appendix}}

  We calculate here the accurate profile of the damping wing of the
Gunn-Peterson trough caused by a homogeneous neutral IGM. The scattering
cross section of the \lya resonance line by neutral hydrogen is given by
(see Peebles 1993, \S 23)

\begin{equation}
\sigma (\omega) =  { 3 \lambda_{\alpha}^2 \Lambda^2 \over 8\pi} \,
{ ( \omega / \omega_{\alpha} )^4 \over
( \omega - \omega_{\alpha} )^2 + 
(\Lambda^2/ 4) \, ( \omega / \omega_{\alpha} )^6 } ~.
\end{equation}
The notation is the same as that used in \S 2, and $\omega_{\alpha} =
2\pi \nu_{\alpha} = 2\pi c / \lambda_{\alpha}$. The second term in the
denominator can be neglected, because we are only interested in the
shape of the damped profile far from the center of the line (i.e.,
when $\| \omega - \omega_{\alpha} \|  \gg \Lambda$, where the resulting
optical depth is not very large. The term in the numerator causes the
classical Rayleigh scattering, and was neglected in \S 2 assuming
$\| \omega - \omega_{\alpha} \| \ll \omega_{\alpha}$, but this is a poor
approximation since the damping wing can be quite broad, and we keep
this term here. We assume that the IGM has a constant neutral hydrogen
comoving density $n_0$ at redshifts $z_n < z < z_s$, where $z_s$ is the
redshift of the source, and the density is negligibly small at
$z < z_n$. The optical depth observed at a wavelength
$\lambda = \lambda_{\alpha} (1+z_s) + \Delta\lambda$ is
\begin{equation}
\tau (\Delta\lambda) = \int_{z_n}^{z_s} { dz\over 1+z} \, 
{c\over H(z)}\, n_0 (1+z)^3 ~ ~
\sigma \left( {\omega \over \omega_{\alpha} } = 
{ (1+z)\over (1+z_s) (1+\delta)} \right) ~,
\end{equation}
where $\delta \equiv \Delta\lambda/[\lambda_\alpha (1+z_s) ] $.
The Gunn-Peterson optical depth outside the damping wing is
$\tau_0(z_s) = 3 \lambda_{\alpha}^3 \Lambda n_0 / [8 \pi c H(z_s)]$, 
and using $H(z) \propto (1+z)^{3/2}$, we obtain:
\begin{equation}
\tau (\Delta\lambda) = {\tau_0 R_{\alpha}\over \pi}\,
\int_{z_n}^{z_s} {dz\over 1+z}\, \left( 1+z \over 1+z_s \right)^{11/2}\,
(1+\delta)^{-4}\, \left[ {1+z\over (1+z_s) \, (1+\delta) } - 1 \right]^{-2} ~.
\end{equation}
This reduces to:
\begin{equation}
\tau (\Delta\lambda) = {\tau_0 R_{\alpha}\over \pi}\,(1+\delta)^{3/2}\,
\int_{x_1}^{x_2} { dx\, x^{9/2} \over (1-x)^2 } ~,
\end{equation}
where $x_1 = (1+z_n)/ [(1+z_s)(1+\delta)]$, $x_2 = (1+\delta)^{-1}$,
and the result of the integral is
\begin{equation}
\int {dx\, x^{9/2}\over (1-x)^2 } = {x^{9/2}\over 1-x} +
{9\over 7} x^{7/2} + {9\over 5} x^{5/2} + 3 x^{3/2} + 9 x^{1/2} -
{9\over 2} \log {1+x^{1/2} \over 1-x^{1/2} } ~.
\end{equation}

\newpage

\begin{figure}
\centerline{
\hbox{
\epsfxsize=4.4truein
\epsfbox[55 32 525 706]{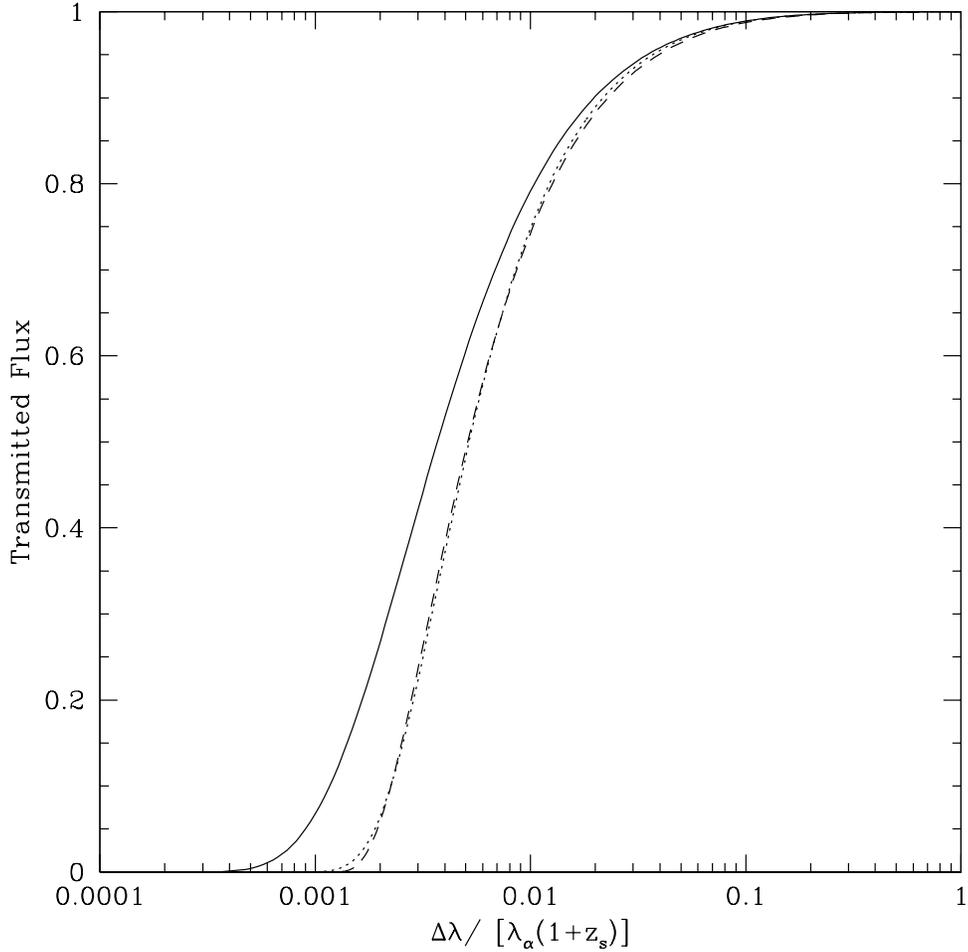}
}
}
\vskip -40pt
\caption{
Absorption profile of the damping wing of the Gunn-Peterson trough,
predicted using the equation given in the Appendix. The fraction of
transmitted flux is shown as a function of the wavelength interval
$\Delta\lambda$ from the \lya wavelength of the source,
$\lambda_{\alpha} (1+z_s)$. The solid line assumes the parameters
$z_s=9$, $\tau_0=4.3\times 10^5$, and $z_n=7$ (where the IGM is assumed
to have a constant neutral density at $z_n < z < z_s$, and be fully
ionized at $z < z_n$). The dotted line is the case where an absorption
system with $\nhi = 2\times 10^{20} \cm^{-2}$ is present at $z_s=9$,
which could be due to a galaxy where the source is located. The dashed
line shows the damping wing profile without any additional absorption
system, but with the parameters changed to $z_s=9.009$ (although
$\Delta\lambda$ is still defined as the wavelength interval from the
\lya line at $z=9$), $\tau_0=4.9\times 10^5$, and provides the best fit
to the dotted line assuming there is no absorption system at the source
redshift. This illustrates the difficulty in measuring $\tau_0$ in the
presence of an absorption system.
}
\end{figure}
\vfill\eject

\end{document}